\documentclass[11pt,titlepage,a4paper]{article}
\usepackage{bozhomac}  % Bozho's style package; includes AMSLaTeX and AMSFonts
\usepackage{bozhlogo}  % the logo of BOZHO commands \BOZHO and \bozho
\usepackage{cite}   % for condensing sequential cited works
\usepackage{varioref}   % flexible cross-references

\title{\bfseries    \vspace*{-1.7in}
{\huge Momentum picture of motion\\[1.2ex]
                in Lagrangian quantum field theory}
}

\vspace{1.7ex}

\author{
Bozhidar Z.\ Iliev
\thanks{Laboratory of Mathematical Modeling in Physics,
Institute for Nuclear Research and \mbox{Nuclear} Energy,
Bulgarian Academy of Sciences,
Boul.\ Tzarigradsko chauss\'ee~72, 1784 Sofia, Bulgaria}
\thanks{E-mail address: bozho@inrne.bas.bg}
\thanks{URL: http://theo.inrne.bas.bg/$\sim$bozho/}
}

\date{  % BEGINNING of \date
 \vspace{2.27ex}\ShortTitle{Momentum picture of motion in QFT}\\[0.27ex]
 \vspace{3.27ex}
\small
    \begin{tabular}{r@{$\colon\to~$}l}
 \vspace{0.09ex} Basic ideas    & June \&\ December, 2001\\[0.09ex]
 \vspace{0.09ex} Began      & January 4, 2002   \\[0.09ex]
 \vspace{0.09ex} Ended      & January 6, 2002   \\[0.09ex]
 \vspace{0.09ex} Initial typeset& January 9 -- January 16, 2002
                            \\[0.09ex]
 \vspace{0.09ex} Last update    & November 1, 2003  \\[0.09ex]
 \vspace{0.27ex} Produced   & \fbox{\today} \\[0.27ex]
    \end{tabular} \\[1.27ex]
\normalsize
    \begin{tabular}{r@{$\colon~$}l}
\vspace{0.27ex} http://www.arXiv.org e-Print archive No. & hep-th/0311003
    \end{tabular} \\[-0.27ex]
 \vspace{4.27ex}{\Huge\BOZHO}   \\[4.27ex]
 \vspace{0.27ex}\Subject{Quantum field theory}
                                \\[2.27ex]
    \begin{tabular}{r@{\hspace{0.512em}}|@{\hspace{0.512em}}l}
 \vspace{0.27ex}\MSC[2000]{81Q99, 81T99\\\hspace{0pt}}% \\[0.27ex]
&
 \vspace{0.27ex}%    \\[0.27ex]
    \end{tabular} \\[1.27ex]
 \vspace{0.27ex}\\[0.27ex]
}%  END of \date{}

\newcommand{\bk}{\boldsymbol{k}}    % Bold "k", 3-dimensional part of
 \newcommand{\Hil}{\mathcal{F}}     % usual Hilbert space ----->>>
    \newcommand{\base}{\mathit{M}}  % Hilbert bundle base space

\newcommand{\ope}[2][{}]{\lindex[\mathcal{#2}]{}{#1}} % operator with left
\newcommand{\tope}[2][{}]{\ope[#1]{\Tilde{#2}}} % operator with Tilde
\begin{document}        % BEGINNING OF THE DOCUMENT

\renewcommand{\thepage}{\roman{page}}

\renewcommand{\thefootnote}{\fnsymbol{footnote}} % special footnote symbols
\maketitle              % the title (page) is put here
\renewcommand{\thefootnote}{\arabic{footnote}}   % usual footnote symbols

\tableofcontents        % the table of contents is put here

%%%%%%%%%%%%%%%%%%%%%%%%%%%%%%%%%%%%%%%%%%%%%%%%%%%%%%%%%%%%%%%%%%%%%%%%%%%%%
%%%%%                                   %%%%%
%%%%%       actual beginning of the document            %%%%%
%%%%%                                   %%%%%
%%%%%%%%%%%%%%%%%%%%%%%%%%%%%%%%%%%%%%%%%%%%%%%%%%%%%%%%%%%%%%%%%%%%%%%%%%%%%

%-->-->-->-->-->-->-->-->-->-->-->-->-->-->-->-->-->-->-->-->-->---->-->-->>
\begin{abstract}

    The basic aspects of the momentum picture of motion in Lagrangian
quantum field theory are given. Under some assumptions, this picture is a
4\ndash dimensional analogue of the Schr{\"o}dinger picture: in it the field
operators are constant, spacetime\ndash independent, while the state vectors
have a simple, exponential, spacetime\ndash dependence. The role of these
assumptions is analyzed. The Euler\ndash Lagrange equations in momentum
picture are derived and attention is paid on the conserved operators in it.

\end{abstract}
%<<--<--<--<--<--<--<--<--<--<--<--<--<--<--<--<--<--<--<--<--<--<--<--<--<--<

\renewcommand{\thepage}{\arabic{page}}

%-->-->-->-->-->-->-->-->-->-->-->-->-->-->-->-->-->-->-->-->-->---->-->-->>
\section {Introduction}
\label{Introduction}
%           BEGINNING OF SECTION~\ref{Introduction}

    The main item of the present  work is a presentation of the basic
aspects
of the \emph{momentum picture of motion} in Lagrangian quantum field theory,
suggested in~\cite{bp-QFT-pictures}. In a sense, under some assumptions, this
picture is a 4\ndash dimensional analogue of the Schr{\"o}dinger picture: in it
the field operators are constant, spacetime\ndash independent, while the
state vectors have a simple, exponential, spacetime\ndash dependence. This
state of affairs offers the known merits of the Schr{\"o}dinger picture (with
respect to Heisenberg one in quantum mechanics~\cite{Messiah-QM}) in the
region of quantum field theory.

    We should mention, in this paper it is considered only the Lagrangian
(canonical) quantum field theory in which the quantum fields are represented
as operators, called field operators, acting on some Hilbert space, which in
general is unknown if interacting fields are studied. These operators are
supposed to satisfy some equations of motion, from them are constructed
conserved quantities satisfying conservation laws, etc. From the view\ndash
point of present\ndash day quantum field theory, this approach is only a
preliminary stage for more or less rigorous formulation of the theory in
which the fields are represented via operator\ndash valued distributions, a
fact required even for description of free fields. Moreover, in non\ndash
perturbative directions, like constructive and conformal field theories, the
main objects are the vacuum mean (expectation) values of the fields and from
these are reconstructed the Hilbert space of states and the acting on it
fields. Regardless of these facts, the Lagrangian (canonical) quantum field
theory is an inherent component of the most of the ways of presentation of
quantum field theory adopted explicitly or implicitly in books
like~\cite{Bogolyubov&Shirkov,Bjorken&Drell,Roman-QFT,Ryder-QFT,
Akhiezer&Berestetskii,Ramond-FT,Bogolyubov&et_al.-AxQFT,Bogolyubov&et_al.-QFT}.
Besides the Lagrangian approach is a source of many ideas for other
directions of research, like the axiomatic quantum field
theory~\cite{Roman-QFT,Bogolyubov&et_al.-AxQFT,Bogolyubov&et_al.-QFT}.

    In Sect.~\ref{Sect2} are reviewed some basic moments of the
Lagrangian formalism in quantum field theory. In Sect~\ref{Sect3} are recalled
part of the relations arising from the assumption that the conserved
operators are generators of the corresponding invariance transformations of
the action integral; in particular the Heisenberg relations between the field
operators and momentum operator are written.

    The momentum picture of motion is defined in Sect~\ref{Sect4}. Two
basic restrictions on the considered quantum field theories is shown to play
a crucial role for the convenience of that picture: the mutual commutativity
between the components of the momentum operator and the Heisenberg
commutation relation between them and the field operators. If these
conditions hold, the field operators in momentum picture become
spacetime\ndash independent and the state vectors turn to have exponential
spacetime\ndash dependence. In Sect~\ref{Sect5}, the attention is called to
the Euler\ndash Lagrange equations and dynamical variables in momentum
picture. In Sect.~\ref{Sect6} is given an idea of the momentum representation
in momentum picture and the similarity with that representation in Heisenberg
picture is pointed. In Sect.~\ref{Sect7} is made a comparison between the
momentum picture in qunatum field theory and the Schr\"odinger pinctur in
quantum mechanics. Some closing remarks are given in Sect.~\ref{Conclusion}.
It is pointed that the above\ndash mentioned restrictions are fundamental
enough to be put in the basic postulates of quantum field theory, which may
result in a new way of its (Lagrangian) construction.

\vspace{1ex}

    The books~\cite{Bogolyubov&Shirkov,Roman-QFT,Bjorken&Drell} will be
used as standard reference works on quantum field theory. Of course, this is
more or less a random selection between the great number of (text)books and
papers on the theme to which the reader is referred for more details or other
points of view. For this end, e.g.,~\cite{Itzykson&Zuber,Ryder-QFT} or the
literature cited
in~\cite{Bogolyubov&Shirkov,Roman-QFT,Bjorken&Drell,Itzykson&Zuber,Ryder-QFT}
may be helpful.

    Throughout this paper $\hbar$ denotes the Planck's constant (divided
by $2\pi$), $c$ is the velocity of light in vacuum, and $\iu$ stands for the
imaginary unit. The superscript $\dag$ means Hermitian conjugation (of
operators or matrices), and the symbol $\circ$ denotes compositions of
mappings/operators.

    The Minkowski spacetime is denoted by $\base$. The Greek indices run
from 0 to $\dim\base=4$. All Greek indices will be raised and lowered by
means of the standard 4\ndash dimensional Lorentz metric tensor
$\eta^{\mu\nu}$ and its inverse $\eta_{\mu\nu}$ with signature
$(+\,-\,-\,-)$. The Einstein's summation convention over indices repeated on
different levels is assumed over the whole range of their values.

%           END OF SECTION~\ref{Introduction}
%<<--<--<--<--<--<--<--<--<--<--<--<--<--<--<--<--<--<--<--<--<--<--<--<--<--<

%-->-->-->-->-->-->-->-->-->-->-->-->-->-->-->-->-->-->-->-->-->-->-->-->-->>
%           BEGINNING OF SECTION~\ref{Sect2}
\section{Lagrangian formalism}
    \label{Sect2}

    Let us consider a system of quantum fields, represented in Heisenberg
picture of motion by field operators $\tope{\varphi}_i(x)\colon\Hil\to\Hil$,
with $i=1,\dots,n\in\field[N]$, in system's Hilbert space $\Hil$ of states
and depending on a point $x$ in Minkowski spacetime $\base$. Here and
henceforth, all quantities in Heisenberg picture, in which the state vectors
are spacetime\ndash independent contrary to the field operators and
observables, will be marked by a tilde (wave)
``$\tope{\mspace{6mu}}\mspace{3mu}$'' over their kernel symbols. Let
    \begin{equation}    \label{2-1}
\tope{L}
= \tope{L}( \tope{\varphi}_i(x), \pd_\nu\tope{\varphi}_j(x) )
    \end{equation}
be the system's Lagrangian, which is supposed to depend on the field
operators and their first partial derivatives.%
\footnote{~%
One can easily generalize the below presented material for Lagrangians
depending on higher order derivatives.%
}
 We expect that this dependence is polynomial or in a form of convergent
power series, which can be treated term by term. The Euler\ndash Lagrange
equations for the Lagrangian~\eref{2-1}, \ie
    \begin{equation}    \label{2-2}
\frac{\pd \tope{L}( \tope{\varphi}_j(x), \pd_\nu\tope{\varphi}_l(x) )}
     {\pd\tope{\varphi}_i(x)}
-
\frac{\pd}{\pd x^\mu}
\frac{\pd \tope{L}( \tope{\varphi}_j(x), \pd_\nu\tope{\varphi}_l(x) )}
     {\pd( \pd_\mu\tope{\varphi}_i(x))}
= 0 ,
    \end{equation}
are identified with the field equations (of motion) for the quantum fields
$\tope{\varphi}_i(x)$.%
\footnote{~%
In~\eref{2-2} and similar expressions appearing further, the derivatives
of functions of operators with respect to operator arguments are calculated
in the same way as if the operators were ordinary (classical)
fields/functions, only the order of the arguments should not be changed.
This is a silently accepted practice in the
literature~\cite{Bogolyubov&Shirkov,Roman-QFT,Bjorken&Drell}. In the most
cases such a procedure is harmless, but it leads to the problem of non\ndash
unique definitions of the quantum analogues of the classical conserved
quantities, like the energy\ndash momentum and charge operators. For some
details on this range of problems in quantum field theory,
see~\cite{bp-QFT-action-principle}. In \emph{loc.\ cit.}\ is demonstrated
that these problems can be eliminated by changing the rules of
differentiation with respect to  \emph{not}\ndash commuting variables. The
paper~\cite{bp-QFT-action-principle} contains an example of a Lagrangian
(describing spin~$\frac{1}{2}$ field) whose field equations are \emph{not}
the Euler\ndash Lagrange equations~\eref{2-2} obtained as just described, but
we shall not investigate such cases in the present work.%
}

    For definiteness, above and below, we consider a quantum field theory
\emph{before} normal ordering and, possibly, without (anti)commutation
relations (see Sect.~\ref{Sect4}). However, our investigation is, practically,
independent of these procedures and can easily be modified to include them.

    Following the standard
procedure~\cite{Bogolyubov&Shirkov,Bjorken&Drell,Itzykson&Zuber,Roman-QFT}
(see also~\cite{bp-QFT-action-principle}), from the Lagrangian~\eref{2-1} can
be constructed the densities of the conserved quantities of the system, \viz
the energy\ndash momentum tensor $\tope{T}_{\mu\nu}(x)$, charge current
$\tope{J}_{\mu}(x)$, the (total) angular momentum density operator
    \begin{equation}    \label{2-3}
\tope{M}_{\mu\nu}^\lambda
= \tope{L}_{\mu\nu}^\lambda(x) + \tope{S}_{\mu\nu}^\lambda(x) ,
    \end{equation}
where
    \begin{equation}    \label{2-4}
\tope{L}_{\mu\nu}^\lambda(x)
= x_\mu \Sprindex[\tope{T}]{\nu}{\lambda}(x)
- x_\nu \Sprindex[\tope{T}]{\mu}{\lambda}(x)
    \end{equation}
and $\tope{S}_{\mu\nu}^\lambda(x)$ are respectively the orbital and spin
angular momentum density operators, and others, if such ones exist. The
corresponding to these quantities integral ones, \viz the momentum, charge,
(total) angular momentum, orbital and spin angular momentum operators, are
respectively defined by:
    \begin{gather}
            \label{2.0}
\tope{P}_\mu
:=
\frac{1}{c}\int\limits_{x^0=\const} \tope{T}_{0\mu}(x) \Id^3\bs x .
\\
            \label{2.10}
\tope{Q} := \frac{1}{c} \int\limits_{x^0=\const} \tope{J}_0(x) \Id^3\bs x
\\          \label{2.11}
\tope{M}_{\mu\nu} = \tope{L}_{\mu\nu}(x) + \tope{S}_{\mu\nu}(x) ,
    \end{gather}
\vspace{-4ex}
    \begin{subequations}    \label{2.12}
      \begin{align}
            \label{2.12a}
\tope{L}_{\mu\nu}(x)
& :=
\frac{1}{c} \int\limits_{x^0=\const}
\{
  x_\mu \Sprindex[\tope{T}]{\nu}{0}(x) - x_\nu \Sprindex[\tope{T}]{\mu}{0}(x)
\} \Id^3\bs x
\\          \label{2.12b}
\tope{S}_{\mu\nu}(x)
& := \frac{1}{c} \int\limits_{x^0=\const} \tope{S}_{\mu\nu}^0(x) \Id^3\bs x
      \end{align}
    \end{subequations}
and satisfy the conservation laws
    \begin{gather}
            \label{2-5}
\frac{\od\tope{P}_\mu}{\od x^0} = 0     \quad
\frac{\od\tope{Q}}{\od x^0} = 0     \quad
\frac{\od\tope{M}_{\mu\nu}}{\od x^0} = 0
\\\intertext{which, in view of~\eref{2.0}--\eref{2.12}, are equivalent to}
            \label{2-6}
\pd_\lambda \tope{P}_\mu = 0    \quad
\pd_\lambda \tope{Q} = 0    \quad
\pd_\lambda \tope{M}_{\mu\nu} = 0
\\\intertext{and also to}
            \label{2-7}
\pd^\lambda \tope{T}_{\lambda\mu} = 0   \quad
\pd^\lambda \tope{J}_\lambda = 0    \quad
\pd_\lambda \tope{M}_{\mu\nu}^\lambda = 0   .
    \end{gather}

    The Lagrangian, as well as the conserved quantities and there
densities, are Hermitian operators; in particular, such is the momentum
operator,
    \begin{equation}    \label{2-8}
\tope{P}_\mu^\dag = \tope{P}_\mu .
    \end{equation}

    The reader can find further details on the Lagrangian formalism in,
e.g.,~\cite{Bogolyubov&Shirkov,Bjorken&Drell,Itzykson&Zuber,Roman-QFT,
bp-QFT-action-principle}.

%           END OF SECTION~\ref{Sect2}
%<<--<--<--<--<--<--<--<--<--<--<--<--<--<--<--<--<--<--<--<--<--<--<--<--<--<

%-->-->-->-->-->-->-->-->-->-->-->-->-->-->-->-->-->-->-->-->-->-->-->-->-->>
\section{Heisenberg relations}
\label{Sect3}
%           BEGINNING OF SECTION~\ref{Sect3}

    The conserved quantities~\eref{2.0}--\eref{2.11} are often identified
with the generators of the corresponding transformations, under which the
action operator is
invariant~\cite{Bogolyubov&Shirkov,Itzykson&Zuber,Bjorken&Drell-2,Roman-QFT}.
This leads to a number of commutation relations between the conserved
operators and between them and the field operators. The relations of the
latter set are often referred as the Heisenberg relations or equations. Part
of them are briefly reviewed below; for details, see \emph{loc.\ cit.}

    The consideration of $\tope{P}_{\mu}$, $\tope{Q}$ and
$\tope{M}_{\mu\nu}$ as generators of translations, constant phase
transformations and 4\ndash rotations, respectively, leads to the following
relations:
    \begin{align}
            \label{3.1}
 [\tope{\varphi}_i(x), \tope{P}_{\mu}]_{\_}
    &= \ih \frac{\pd \tope{\varphi}_i(x)}{\pd x^\mu}
\\
            \label{2.17}
 [\tope{\varphi}_i(x), \tope{Q}]_{\_}
    &= \varepsilon(\tope{\varphi}_i) q_i \tope{\varphi}_i(x)
\\          \label{2.18}
 [\tope{\varphi}_i(x), \tope{M}_{\mu\nu}]_{\_}
&=
\ih\{
x_\mu\pd_\nu\tope{\varphi}_i(x) - x_\nu\pd_\mu\tope{\varphi}_i(x)
+ I_{i\mu\nu}^{j} \tope{\varphi}_j(x)
\} .
    \end{align}
Here: $q_i=\const$ is the charge of the $i^{\text{th}}$ field,
$\varepsilon(\tope{\varphi}_i) = 0$ if
        $\tope{\varphi}_i^\dag = \tope{\varphi}_i$,
$\varepsilon(\tope{\varphi}_i) = \pm 1$ if
        $\tope{\varphi}_i^\dag \not= \tope{\varphi}_i$
with
$\varepsilon(\tope{\varphi}_i) + \varepsilon(\tope{\varphi}_i^\dag) = 0$,
and the constants $I_{i\mu\nu}^{j} = -I_{i\nu\mu}^{j}$ characterize the
transformation properties of the field operators under 4\ndash rotations.
(It is a convention whether to put $\varepsilon(\tope{\varphi}_i)=+1$
or $\varepsilon(\tope{\varphi}_i)=-1$ for a fixed $i$.)
    Besides, the operators $\tope{P}_{\mu}$, $\tope{Q}$ and
$\tope{M}_{\mu\nu}$ satisfy certain commutation relation between themselves,
from which we shall write the following two:
    \begin{align}   \label{2.1}
 [\tope{P}_\mu, \tope{P}_\nu ]_{\_} & = 0
\\          \label{2.20}
 [\tope{Q}, \tope{P}_\mu ]_{\_} & = 0 .
    \end{align}

    It should be clearly understood, the equations~\eref{3.1}--\eref{2.20}
are from pure geometrical origin and are completely external to the
Lagrangian formalism. However, there are strong evidences that they should
hold in a realistic Lagrangian quantum field theory
(see~\cite[\S~68]{Bjorken&Drell-2}
and~\cite[\S~5.3 and~\S~9.4]{Bogolyubov&Shirkov}). Moreover, (most of) the
above relations happen to be valid for Lagrangians that are frequently used,
\eg for the ones describing free fields~\cite{Bjorken&Drell-2}.

%           END OF SECTION~\ref{Sect3}
%<<--<--<--<--<--<--<--<--<--<--<--<--<--<--<--<--<--<--<--<--<--<--<--<--<--<

%-->-->-->-->-->-->-->-->-->-->-->-->-->-->-->-->-->-->-->-->-->-->-->-->-->>
\section{The momentum picture of motion}
\label{Sect4}
%           BEGINNING OF SECTION~\ref{Sect4}

    Let $\tope{P}_{\mu}$ be the system's momentum operator, given
by equation~\eref{2.0}. Since $\tope{P}_{\mu}$ is Hermitian (see~\eref{2-8}),
the operator
    \begin{equation}    \label{12.112}
\ope{U}(x,x_0)
 =
\exp\Bigl( \iih \sum_\mu (x^\mu-x_0^\mu)\tope{P}_{\mu}  \Bigr) ,
    \end{equation}
where $x_0\in\base$ is arbitrarily fixed and $x\in\base$, is unitary, \ie
    \begin{equation}    \label{4.1}
\ope{U}^\dag(x_0,x) := (\ope{U}(x,x_0))^\dag
 =
(\ope{U}(x,x_0))^{-1}  =: \ope{U}^{-1}(x,x_0) .
    \end{equation}

    Let $\tope{X}\in\Hil$ be a state vector in the system's Hilbert space
$\Hil$ and $\tope{A}(x)\colon\Hil\to\Hil$ be an operator on it. The
transformations
    \begin{align}   \label{12.113}
\tope{X}\mapsto \ope{X}(x)
    &= \ope{U}(x,x_0) (\tope{X})
\\          \label{12.114}
\tope{A}(x)\mapsto \ope{A}(x)
    &= \ope{U}(x,x_0)\circ (\tope{A}(x)) \circ \ope{U}^{-1}(x,x_0) ,
    \end{align}
evidently, preserve the Hermitian scalar product
 $\langle\cdot | \cdot\rangle \colon \Hil\times\Hil \to \field[C]$ of $\Hil$
and the mean values of the operators, \ie
    \begin{equation}    \label{4.2}
\langle \tope{X} | \tope{A}(x)(\tope{Y}) \rangle
=
\langle \ope{X}(x) | \ope{A}(x)(\ope{Y}(x)) \rangle
    \end{equation}
for any $\tope{X},\tope{Y}\in\Hil$ and $\tope{A}(x)\colon\Hil\to\Hil$. Since
the physically predictable/measurable results of the theory are expressible
via scalar products in
$\Hil$~\cite{Bogolyubov&Shirkov,Bjorken&Drell,Itzykson&Zuber}, the last
equality implies that the theory's description via vectors and operators
like $\tope{X}$ and $\tope{A}(x)$ above is completely equivalent to the one
via the vectors $\ope{X}(x)$ and operators $\tope{A}(x)$, respectively. The
description of quantum field theory via $\ope{X}$ and $\ope{A}(x)$ will be
called the \emph{momentum picture (of motion (of quantum field
theory))}~\cite{bp-QFT-pictures}.

    However, without further assumptions, this picture turns to be rather
complicated. The mathematical cause for this is that derivatives of
different operators are often met in the theory and, as a consequence
of~\eref{12.114}, they transform as
    \begin{gather}  \label{4.3}
\pd_\mu \tope{A}(x) \mapsto
\ope{U}(x,x_0) \circ (\pd_\mu \tope{A}(x)) \circ \ope{U}^{-1}(x,x_0)
=
\pd_\mu \ope{A}(x) + [\ope{A}(x), \ope{H}_\mu(x,x_0) ]_{\_}
\\          \label{4.4}
\ope{H}_\mu(x,x_0)
:= \bigl( \pd_\mu \ope{U}(x,x_0) \bigr) \circ \ope{U}^{-1}(x,x_0)
    \end{gather}
from Heisenberg to momentum picture. Here
$[\ope{A},\ope{B}]_{\_}:=\ope{A}\circ\ope{B}-\ope{B}\circ\ope{A}$ is the
commutator of $\ope{A},\ope{B}\colon\Hil\to\Hil$. The entering in~\eref{4.3},
via~\eref{4.4}, derivatives of the operator~\eref{12.112} can be represented
as the convergent power series
\[
\pd_\mu \ope{U}(x,x_0)
=
\iih \ope{P}_\mu
+ \iih \sum_{n=1}^{\infty} \frac{1}{(\ih)^n} \frac{1}{(n+1)!} \sum_{m=0}^{n}
\bigl( (x^\lambda-x_0^\lambda)\tope{P}_\lambda \bigr)^m
   \circ \tope{P}_\mu \circ
\bigl( (x^\lambda-x_0^\lambda)\tope{P}_\lambda \bigr)^{n-m},
\]
where $(\cdots)^{n}:=(\cdots)\circ\cdots\circ(\cdots)$ ($n$\ndash times) and
$(\cdots)^0:=\id_\Hil$ is the identity mapping of $\Hil$, which cannot be
written in a closed form unless the commutator
$[\tope{P}_\mu,\tope{P}_\nu]_{\_}$ has `sufficiently simple' form. In
particular, the relation~\eref{2.1} entails
 $\pd_\mu \ope{U}(x,x_0) = \iih \tope{P}_\mu\circ\ope{U}(x,x_0)$, so
that~\eref{4.4} and~\eref{4.3} take respectively the form
    \begin{gather}  \label{4.5}
\ope{H}_\mu(x,x_0) = \iih \tope{P}_\mu
\\          \label{4.6}
\pd_\mu \tope{A}(x)
\mapsto
\pd_\mu \ope{A}(x) + \iih [\ope{A}(x), \tope{P}_\mu ]_{\_} .
    \end{gather}
Notice, the equality~\eref{4.5} is possible iff and only if
$\ope{U}_\mu(x,x_0)$ is a solution of the initial\ndash value problem
(see~\eref{4.4})
    \begin{subequations}    \label{12.116}
    \begin{align}   \label{12.116a}
\ih \frac{\pd\ope{U}(x,x_0)}{\pd x^\mu}
         & = \tope{P}_{\mu} \circ \ope{U}(x,x_0)
\\                      \label{12.116b}
\ope{U}(x_0,x_0) &= \id_\Hil ,
    \end{align}
    \end{subequations}
the integrability conditions for which are exactly~\eref{2.1}.%
\footnote{~\label{fn:momentum-curvature}%
For a system with a non-conserved momentum operator $\tope{P}_\mu(x)$ the
operator $\ope{U}(x,x_0)$ should be defined as the solution of~\eref{12.116},
with $\tope{P}_\mu(x)$ for $\tope{P}_\mu$, instead of by~\eref{12.112}; in
this case, equation~\eref{2.1} should be replace with
\[
[ \tope{P}_\mu(x) , \tope{P}_\nu(x) ]_{\_}
        + \pd_\nu \tope{P}_\mu(x) - \pd_\mu \tope{P}_\nu(x) = 0 .
\]
Most of the material in the present section remains valid in that more
general situation.%
}
Since~\eref{2.1} and~\eref{12.112} imply
    \begin{equation}    \label{2.2}
[  \ope{U}(x,x_0) , \tope{P}_\mu ]_{\_} =0,
    \end{equation}
by virtue of~\eref{12.114}, we have
    \begin{equation}    \label{2.3}
     \ope{P}_\mu = \tope{P}_\mu,
    \end{equation}
\ie the momentum operators in Heisenberg and momentum pictures coincide,
provided~\eref{3.1} holds.

    It is worth to be mentioned, equation~\eref{2.3} is a special case of
    \begin{gather}  \label{4.8}
\ope{A}(x)
= \tope{A}(x)
+ [\ope{U}(x,x_0), \tope{A}(x)]_{\_} \circ \ope{U}^{-1}(x,x_0) ,
\\\intertext{which is a consequence of~\eref{12.114} and is quite useful if
one knows explicitly the commutator $[\ope{U}(x,x_0), \tope{A}(x)]_{\_}$. In
particular, if}
            \label{4.9}
\bigl[ [\tope{A}(x), \tope{P}_\mu]_{\_} , \tope{P}_\nu \bigr]_{\_} = 0
\\\intertext{and~\eref{2.1} holds, then, by expanding~\eref{12.112} into a
power series, one can prove that}
            \label{4.10}
[\tope{A}(x),\ope{U}(x,x_0)]_{\_}
=
\iih (x^\lambda-x_0^\lambda)
     [\tope{A}(x), \tope{P}_\lambda]_{\_} \circ \ope{U}(x,x_0) .
\\\intertext{So, in this case,~\eref{4.8} reduces to}
            \label{4.11}
\ope{A}(x)
=
  \tope{A}(x)
- \iih (x^\lambda-x_0^\lambda) [\tope{A}(x),\tope{P}_\lambda]_{\_} .
    \end{gather}
This formula allows to be found an operator in momentum picture if its
commutator(s) with (the components of) the momentum operator is (are)
explicitly known, provided~\eref{2.1} and~\eref{4.9} hold. The choice
$\tope{A}(x)=\tope{P}_\mu$ reduces~\eref{4.11} to~\eref{2.3}.

    Of course, a transition from one picture of motion to other one is
justified if there are some merits from this step; for instance, if some
(mathematical) simplification, new physical interpretation etc.\ occur in the
new picture. A classical example of this kind is the transition between
Schr{\"o}dinger and Heisenberg pictures in quantum mechanics~\cite{Messiah-QM}
or, in a smaller extend, in quantum field theory~\cite{Bogolyubov&Shirkov}.
Until now we have not present evidences that the momentum picture can bring
some merits with respect to, e.g., Heisenberg picture. On the opposite, there
was an argument that, without further restrictions, mathematical
complications may arise in it.

    In this connection, let us consider, as a second possible
restriction, a theory in which the Heisenberg equation~\eref{3.1} is valid.
In momentum picture, it reads
    \begin{equation}    \label{4.12}
[ \varphi_i(x), \tope{P}_\mu - \ih \ope{H}_\mu ]_{\_}
=
 \ih \pd_\mu \varphi_i(x),
    \end{equation}
where
    \begin{align}   \label{12.115}
\tope{\varphi}_i(x)\mapsto \ope{\varphi}_i(x)
     = \ope{U}(x,x_0)\circ \tope{\varphi}_i(x) \circ \ope{U}^{-1}(x,x_0) .
    \end{align}
are the field operators in momentum picture and the relations~\eref{12.114}
and~\eref{4.3} were applied. The equation~\eref{4.12} shows that,
if~\eref{4.5} holds, which is equivalent to the validity of~\eref{2.1}, then
    \begin{equation}    \label{4.13}
\pd_\mu \varphi_i(x) = 0 ,
    \end{equation}
\ie in this case the field operators in momentum picture turn to be constant,
    \begin{equation}    \label{2.4}
\varphi_i(x)
= \ope{U}(x,x_0)\circ \tope{\varphi}_i(x) \circ \ope{U}^{-1}(x,x_0)
= \varphi_i(x_0)
= \tope{\varphi}_i(x_0)
=: \varphi_{i} .
    \end{equation}
As a result of the last fact, all functions of the field operators and their
derivatives, polynomial or convergent power series in them, become constant
operators in momentum picture, which are algebraic functions of the field
operators in momentum picture. This is an essentially new moment in the
theory that reminds to a similar situation in the Schr{\"o}dinger picture in
quantum mechanics (see~\cite{Messiah-QM} and Sect.~\ref{Sect7} below).

    If $\tope{P}_\mu$ is considered, as
usual~\cite{Bogolyubov&Shirkov,Bjorken&Drell}, as a generator of 4\ndash
translations, then the constancy of the field operators in momentum picture
is quite natural. In fact, in this case, the transition
$\tope{\varphi}_i(x)\mapsto\ope{\varphi}_i(x)$, given by~\eref{12.115}, means
that the argument of $\tope{\varphi}_i(x)$ is shifted by $(x_0-x)$, \ie that
\(
\tope{\varphi}_i(x) \mapsto
\ope{\varphi}_i(x) = \tope{\varphi}_i(x+(x_0-x)) =\tope{\varphi}_i(x_0) .
\)

    Let us turn our attention now to system's state vectors. By
definition~\cite{Bogolyubov&Shirkov,Bjorken&Drell}, such a vector $\tope{X}$
is a spacetime\ndash constant one in Heisenberg picture,
    \begin{equation}    \label{4.14}
\pd_\mu \tope{X} = 0 .
    \end{equation}
In momentum picture, the situation is opposite, as, by virtue
of~\eref{12.113}, the operator~\eref{12.112} plays a role of spacetime
`evolution' operator, \ie
    \begin{equation}    \label{12.118}
\ope{X}(x)
= \ope{U}(x,x_0) (\ope{X}(x_0))
= \e^{\iih(x^\mu-x_0^\mu)\tope{P}_{\mu} } (\ope{X}(x_0)) ,
    \end{equation}
with
    \begin{equation}    \label{4.15}
\ope{X}(x_0) = \ope{X}(x)|_{x=x_0} = \tope{X}
    \end{equation}
being considered as initial value of $\ope{X}(x)$ at $x=x_0$. Thus, if
$\ope{X}(x_0)=\tope{X}$ is an eigenvector of the momentum operators
$\tope{P}_\mu=\ope{P}_\mu(x)|_{x=x_0}$
($=\{ \ope{U}(x,x_0) \circ\tope{P}_\mu\circ \ope{U}^{-1}(x,x_0) \}|_{x=x_0} $)
with eigenvalues $p_\mu$, \ie
    \begin{equation}    \label{4.16}
\tope{P}(\tope{X}) = p_\mu \tope{X}
\qquad
\bigl(\, =p_\mu\ope{X}(x_0) = \ope{P}_\mu(x_0)(\tope{X}(x_0))\, \bigr) ,
    \end{equation}
we have the following \emph{explicit} form of a state vector $\ope{X}$:
    \begin{equation}    \label{12.120}
\ope{X}(x)
=
\e^{ \iih(x^\mu-x_0^\mu)p_\mu } (\ope{X}(x_0)).
    \end{equation}
It should be understood, this is the \emph{general form of all state vectors
in momentum picture}, as they are eigenvectors of all (commuting)
observables~\cite[p.~59]{Roman-QFT}, in particular, of the momentum
operator.

    So, in momentum picture, the state vectors have a relatively simple
\emph{global} description. However, their differential (local) behavior is
described via a differential equation that may turn to be rather complicated
unless some additional conditions are imposed. Indeed, form~\eref{12.118}, we
get
    \begin{equation}    \label{4.17}
\pd_\mu \ope{X}(x) = \ope{H}_\mu(x,x_0) (\ope{X}(x))
    \end{equation}
in which equation the operator $\ope{H}_\mu(x,x_0)$ is given by~\eref{4.4}
and may have a complicated explicit form (\emph{vide supra}). The
equality~\eref{4.17} has a form similar to the one of the Schr{\"o}dinger
equation, but in `4\ndash dimensions', with `4\ndash dimensional Hamiltonian'
$\ih\ope{H}_\mu(x,x_0)$.  It is intuitively clear, in this context, the
operators $\ih\ope{H}_\mu(x,x_0)$ should be identified with the components of
the momentum operator $\ope{P}_\mu$, \ie the equality~\eref{4.5} is a
natural one on this background.

    Thus, if we accept~\eref{4.5}, or equivalently~\eref{2.1}, a state
vector $\ope{X}(x)$ in momentum picture will be a solution of the
initial\ndash value problem
    \begin{equation}    \label{12.117}
\ih \frac{\pd\ope{X}(x)}{\pd x^\mu}
=
\ope{P}_{\mu}  (\ope{X}(x))
\qquad
\ope{X}(x)|_{x=x_0}=\ope{X}(x_0) = \tope{X}
    \end{equation}
and, respectively, the evolution operator $\ope{U}(x,x_0)$ of the state
vectors will be a solution of~\eref{12.116}. Consequently, the
equation~\eref{2.1} entails not only a simplified description of the
operators in momentum picture, but also a natural one of the state vectors in
it.

    The above discussion reveals that the momentum picture is worth to be
employed in quantum field theories in which the conditions
    \begin{subequations}    \label{4.18}
    \begin{align}   \label{4.18a}
& [\tope{P}_\mu, \tope{P}_\nu ]_{\_} = 0
\\          \label{4.18b}
& [\tope{\varphi}_i(x), \tope{P}_\mu]_{\_} = \ih\pd_\mu \tope{\varphi}_i(x)
    \end{align}
    \end{subequations}
are valid. In that case, the momentum picture can be considered as a 4\ndash
dimensional analogue of the Schr{\"o}dinger picture~\cite{bp-QFT-pictures}: the
field operators are spacetime\ndash constant and the state vectors are
spacetime\ndash dependent and evolve according to the `4\ndash dimensional
Schr{\"o}dinger equation'~\eref{12.117} with evolution
operator~\eref{12.112}. More details on that item will be given in
Sect.~\ref{Sect7} below.

    In connection with the conditions~\eref{4.18}, it should be said that
their validity is more a rule than an exception. For instance, in the
axiomatic quantum field theory, they hold identically as in this approach, by
definition, the momentum operator is identified with the generator of
translations~\cite{Bogolyubov&et_al.-AxQFT,Bogolyubov&et_al.-QFT}. In the
Lagrangian formalism, to which~\eref{4.18} are external restrictions, the
conditions~\eref{4.18} seem to hold at least for the investigated free fields
and most (all?) interacting ones~\cite{Bjorken&Drell-2}. For example, the
commutativity between the components of the momentum operator, expressed
via~\eref{4.18a}, is a consequence of the (anti)commutation relations and,
possibly, the field equations. Besides, it expresses the simultaneous
measurability of the components of system's momentum. The Heisenberg
relation~\eref{4.18b} is verified in ~\cite{Bjorken&Drell-2} for a number of
Lagrangians. Moreover, in \emph{loc.\ cit.}\ it is regarded as one of the
conditions for relativistic covariance in a translation\ndash invariant
Lagrangian quantum field theory. All these facts point that the
conditions~\eref{4.18} are fundamental enough to be incorporated in the basic
postulates of quantum field theory, as it is done (more implicitly than
explicitly), e.g., in~\cite{Bogolyubov&Shirkov,Itzykson&Zuber,Roman-QFT}.
Some comments on that problem will be presented in Sect.~\ref{Conclusion}
(see also~\cite[chapter~1]{Ohnuki&Kamefuchi}).

%           END OF SECTION~\ref{Sect4}
%<<--<--<--<--<--<--<--<--<--<--<--<--<--<--<--<--<--<--<--<--<--<--<--<--<--<

%-->-->-->-->-->-->-->-->-->-->-->-->-->-->-->-->-->-->-->-->-->-->-->-->-->>
\section
[General aspects of Lagrangian formalism in momentum picture]
{General aspects of Lagrangian formalism\\ in momentum picture}
\label{Sect5}
%           BEGINNING OF SECTION~\ref{Sect5}

    In this section, some basic moments of the Lagrangian formalism in
momentum picture will be considered, provided the equations~\eref{4.18} hold.

    To begin with, let us recall, in the momentum picture, under the
conditions~\eref{4.18}, the field operators $\varphi_i$ are constant, \ie
spacetime\ndash independent (which is equivalent to~\eref{4.18a}), and the
state vectors are spacetime\ndash dependent, their dependence being of
exponential type (see~\eref{12.118} and~\eref{12.120}). As a result of this,
one can expect a simplification of the formalism, as it happens to be the
case.

    Combining~\eref{4.3}, with
$\ope{A}=\varphi_i$,~\eref{4.5},~\eref{4.13} and~\eref{2.3}, we see that the
first partial derivatives of the field operators transform from Heisenberg to
momentum picture according to the rule
    \begin{equation}    \label{5.1}
\pd_\mu \tope{\varphi}_i(x) \mapsto
y_{i\mu} := \iih [\varphi_i,\ope{P}_\mu]_{\_} .
    \end{equation}
Therefore the operator $\pd_\mu$, when applied to field operators, transforms
into $\iih[\cdot,\tope{P}_\mu]_{\_}=\iih[\cdot,\ope{P}_\mu]_{\_}$, which is a
differentiation of the operator space over $\Hil$. An important corollary
of~\eref{5.1} is that any (finite order) differential expression of
$\tope{\varphi}_i(x)$ transforms in momentum picture into an \emph{algebraic}
one of $\varphi_i$. In particular, this concerns the Lagrangian (which is
supposed to be polynomial or convergent power series in the field operators
and their partial derivatives):
    \begin{align*}
\tope{L} \mapsto \ope{L}
: & = \ope{L}(\varphi_i(x))
:=
\ope{U}(x,x_0)\circ
\tope{L}( \tope{\varphi}_i(x),\pd_\nu\tope{\varphi}_j(x) )
\circ \ope{U}^{-1}(x,x_0)
\\ \notag & =
\tope{L} \bigl(
\ope{U}(x,x_0)\circ  \tope{\varphi}_i(x)  \circ \ope{U}^{-1}(x,x_0)
,
\ope{U}(x,x_0)\circ  \pd_\nu\tope{\varphi}_j(x)  \circ \ope{U}^{-1}(x,x_0)
\bigr)
\\ %\notag
%   \label{12.125}
& =
\tope{L} \bigl( \ope{\varphi}_i , \iih [\varphi_j,\ope{P}_\nu ]_{\_}
\bigr)  .
    \end{align*}
Thus, the Lagrangian~\eref{2-1} in momentum picture reads
    \begin{equation}    \label{2.5}
\ope{L}
= \ope{L} (\phi_i)
= \tope{L}(\varphi_i, y_{j\nu})
\qquad
y_{j\nu}=\iih[\varphi_j,\ope{P}_{\nu} ]_{\_}  ,
    \end{equation}
\ie one has to make simply the replacements
$\tope{\varphi}_i(x)\mapsto\varphi_i$ and
$\pd_\nu\tope{\varphi}_i(x)\mapsto y_{i\nu}$ in~\eref{2-1}.

    Applying the general rule~\eref{12.114} to the Euler-Lagrange
equations~\eref{2-2} and using~\eref{2.4} and~\eref{5.1}, we find, after some
simple calculations,%
\footnote{~%
For details, see~\cite{bp-QFT-pictures}.%
}
the \emph{Euler\ndash Lagrange equations in momentum picture} as
    \begin{equation}    \label{12.129}
\Bigl\{
\frac{\pd\tope{L}(\varphi_j,y_{l\nu})} {\pd \varphi_i}
-
\iih
\Bigl[
\frac{\pd\tope{L}(\varphi_j,y_{l\nu})} {y_{i\mu}} , \ope{P}_{\mu}
\Bigr]_{\_}
\Bigr\}
\Big|_{ y_{j\nu}=\iih[\varphi_j,\ope{P}_{\nu} ]_{\_} }
= 0 .
    \end{equation}
A feature of these equations is that they are \emph{algebraic}, not
differential, ones with respect to the field operators $\varphi_i$ (in
momentum picture), provided $\ope{P}_\mu$ is regarded as a given known
operator. This is a natural fact in view of~\eref{4.13}.

    We shall illustrate the above general considerations on the almost
trivial example of a free Hermitian scalar field $\tope{\varphi}$, described
in Heisenberg picture by the Lagrangian
\(
\tope{L}
= - \frac{1}{2}m^2c^4\tope{\varphi}\circ\tope{\varphi}
  + c^2\hbar^2 (\pd_\mu\tope{\varphi})\circ(\pd^\mu\tope{\varphi})
=\tope{L}(\tope{\varphi},\widetilde{y}_\nu),
\)
with $m=\const$ and  $\widetilde{y}_\nu=\pd_\nu\tope{\varphi}$,
and satisfying the Klein\ndash Gordon equation
$(\widetilde{\square}+\frac{m^2c^2}{\hbar^2} \id_\Hil) \tope{\varphi} = 0 $,
$\widetilde{\square}:=\pd_\mu\pd^\mu$. In momentum picture
$\tope{\varphi}$ transforms into the constant operator

    \begin{equation}    \label{12.130}
\varphi(x)
= \ope{U}(x,x_0)\circ \tope{\varphi} \circ \ope{U}^{-1}(x,x_0)
= \varphi(x_0)
= \tope{\varphi}(x_0)
=: \varphi
    \end{equation}
which, in view of~\eref{12.129},
 $\frac{\pd\tope{L}} {\pd\varphi} = -m^2c^4\varphi$, and
 $\frac{\pd\tope{L}} {\pd y_\nu} = c^2\hbar^2 y_\mu\eta^{\mu\nu} $
is a solution of
    \begin{equation}    \label{12.131}
m^2c^2\varphi
-
[ [
    \varphi,\ope{P}_\mu ]_{\_},\ope{P}^\mu
]_{\_}
= 0 .
    \end{equation}
This is the \emph{Klein-Gordon equation in momentum picture}. As a
consequence of~\eref{2.3}, this equation is valid in Heisenberg picture too,
when it is also a corollary of the Klein\ndash Gordon equation and the
Heisenberg relation~\eref{4.18b}.

    The Euler-Lagrange equations~\eref{12.129} are not enough for
determination of the field operators $\varphi_i$. This is due to the simple
reason that in them enter also the components $\ope{P}_\mu$ of the
(canonical) momentum operator~\eref{2.0}, which are functions (functionals)
of the field operators. Hence, a complete system of equations for the field
operators should consists of~\eref{12.129} and an explicit connection between
them and the momentum operator.

    Since the densities of the conserved operators of a system are
polynomial functions of the field operators and their partial derivatives in
Heisenberg picture (for a polynomial Lagrangian of type~\eref{2-1}), in
momentum picture they became polynomial functions of $\varphi_i$ and
$y_{j\nu}=\iih[\varphi_j,\ope{P}_{\nu} ]_{\_}$.
When working in momentum picture, in view of~\eref{12.114}, the following
representations turn to be useful:
      \begin{gather}
            \label{2.6}
\ope{P}_\mu = \tope{P}_\mu
=
\frac{1}{c} \int\limits_{x^0=\const}
    \ope{U}^{-1}(x,x_0)\circ \ope{T}_{0\mu} \circ \ope{U}(x,x_0)
    \Id^3\bs{x}
\displaybreak[1]\\              \label{2.13}
\tope{Q}
=
\frac{1}{c} \int\limits_{x^0=\const}
    \ope{U}^{-1}(x,x_0)\circ \ope{J}_{0} \circ \ope{U}(x,x_0)
    \Id^3\bs{x}
\displaybreak[1]\\              \label{2.14}
\tope{L}_{\mu\nu} (x)
=
\frac{1}{c} \int\limits_{x^0=\const}
    \ope{U}^{-1}(x,x_0)\circ \{
  x_\mu \Sprindex[\ope{T}]{\nu}{0} - x_\nu \Sprindex[\ope{T}]{\mu}{0}
\} \circ \ope{U}(x,x_0)
    \Id^3\bs{x}
\displaybreak[1]\\              \label{2.15}
\tope{S}_{\mu\nu} (x)
=
\frac{1}{c} \int\limits_{x^0=\const}
    \ope{U}^{-1}(x,x_0)\circ \ope{S}_{\mu\nu}^0 \circ \ope{U}(x,x_0)
    \Id^3\bs{x} .
      \end{gather}
In particular, the combination of~\eref{2.6} and~\eref{12.129} (together with
an explicit expression for the energy\ndash momentum tensor
$\ope{T}_{\mu\nu}$) provide a closed algebraic\ndash functional system of
equations for determination of the field operators $\varphi_i$ in momentum
picture. In fact, this is the \emph{system of field equations in momentum
picture}. Concrete types of such systems of field equations and their
links with the (anti)commutation (and paracommutation) relations will be
investigated elsewhere.

    In principle, from~\eref{2.6}--\eref{2.15} and the field equations
(i.e.~\eref{2.6} and~\eref{12.129}) can be found the commutation relations
between the conserved quantities and the momentum operator, \ie
$[\tope{D},\tope{P}_\lambda]_{\_}$ with
$\tope{D}=\tope{P}_\mu,\tope{Q},\tope{M}_{\mu\nu}$. If one succeeds in
computing $[\tope{D},\tope{P}_\lambda]_{\_}$, one can calculate
$[\tope{D},\ope{U}(x,x_0)]_{\_}$ and, via~\eref{4.8}, the operator
$\ope{D}=\ope{P}_\mu,\ope{Q},\ope{M}_{\mu\nu}$ in momentum picture. If it
happens that~\eref{4.9} holds for $\tope{A}=\tope{D}$, then one can use
simply the formula~\eref{4.11}. In particular, this is the case if the
commutators $[\tope{D},\tope{P}_\lambda]_{\_}$ coincide with relations
like~\eref{2.1} and~\eref{2.20} (see also~\cite{Bjorken&Drell-2,Roman-QFT}.%
\footnote{~%
In future work(s), it will be proved that, in fact, the so-calculated
commutators $[\tope{D},\tope{P}_\lambda]_{\_}$ reproduce similar relations,
obtained from pure geometrical reasons in Heisenberg picture, at least for
the most widely used Lagrangians. However, for the above purpose, one cannot
use directly the last relations, except~\eref{2.1} in this case, because
they are external to the Lagrangian formalism, so that they represent
additional restriction to its consequences.%
}
For instance, if~\eref{2.20} holds, then~\eref{4.11} yields
$\ope{Q}=\tope{Q}$, \ie the charge operator remains one and the same in
momentum and in Heisenberg pictures. Obviously, the last result holds for
any operator commuting with the momentum operator.

    A constant operator $\tope{C}$ in Heisenberg picture,
    \begin{equation}    \label{5.10}
\pd_\mu \tope{C} = 0,
    \end{equation}
transforms in momentum picture into an operator $\ope{C}(x)$ such that
    \begin{equation}    \label{5.11}
\pd_\mu \ope{C}(x) + \iih [\ope{C}(x),\ope{P}_\mu]_{\_} = 0,
    \end{equation}
due to~\eref{4.6} and~\eref{2.3}. In particular, the conserved quantities
(e.g., the momentum, charge and angular momentum operators) are solutions of
equation~\eref{5.11}, \ie a conserve operator need not to be a constant one
in momentum picture, but it necessarily satisfies~\eref{5.11}. Obviously, a
constant operator $\tope{C}$ in Heisenberg picture is such in momentum
picture if and only if it commutes with the momentum operator,
    \begin{equation}    \label{5.12}
\pd_\mu \ope{C}(x) = 0 \iff  [\ope{C}(x),\ope{P}_\mu]_{\_} = 0.
    \end{equation}
Such an operator, by virtue of~\eref{4.9} and~\eref{4.11}, is one and the
same in Heisenberg and momentum pictures,
    \begin{equation}    \label{5.13}
\ope{C}(x) = \tope{C} .
    \end{equation}
In particular, the dynamical variables which are simultaneously measurable
with the momentum, \ie commuting with  $\tope{P}_\mu$, remain
constant in momentum picture and, hence, coincide with their values in
Heisenberg one. Of course, such an operator is $\tope{P}_\mu=\ope{P}_\mu$, as
we suppose the validity of~\eref{4.18a}, and the charge operator
$\tope{Q}=\ope{Q}$, if~\eref{2.20} holds.

    Evidently, equation~\eref{5.11} is a 4-dimensional analogue of
\(
\ih\frac{\pd\ope{A}(t)}{\pd t} + [\ope{A}(t),\ope{H}(t)]_{\_} = 0,
\)
which is a necessary and sufficient condition (in Schr{\"o}dinger picture)
for an observable $\ope{A}(t)$ to be an integral of motion of a quantum
system with Hamiltonian $\ope{H}(t)$ in non\ndash relativistic quantum
mechanics~\cite{Messiah-QM,Dirac-PQM}.

    At the end, let us consider the Heisenberg
relations~\eref{3.1}--\eref{2.18} in momentum picture. As we said above, the
first of them reduces to~\eref{4.13} in momentum picture and simply expresses
the constantcy of the field operators $\varphi_i$. Since~\eref{2.17} has
polynomial structure with respect to $\tope{\varphi}_i(x)$, the transition to
momentum picture preserves it, \ie we have (see~\eref{12.114})
    \begin{equation}    \label{2.26}
[\ope{\varphi}_i, \ope{Q}]_{\_}
= \varepsilon(\ope{\varphi}_i) q_i \ope{\varphi}_i
\qquad
\varepsilon(\ope{\varphi}_i) = \varepsilon(\tope{\varphi}_i) .
    \end{equation}
At last, applying~\eref{12.114} to the both sides of~\eref{2.18} and taking
into account~\eref{5.1}, we obtain
    \begin{equation}    \label{2.27}
[\ope{\varphi}_i, \ope{M}_{\mu\nu}(x,x_0)]_{\_}
=
x_\mu [\varphi_i ,\ope{P}_\nu]_{\_} - x_\nu [\varphi_i ,\ope{P}_\mu]_{\_}
+ \ih I_{i\mu\nu}^{j} \varphi_j .
    \end{equation}
However, in a pure Lagrangian approach, to which~\eref{2.26} and~\eref{2.27}
are external restrictions, one is not allowed to apply~\eref{2.26}
and~\eref{2.27} unless these equations are explicitly proved for the
operators $\ope{M}_{\mu\nu}$ and $\ope{P}_{\mu}$ given
via~\eref{2.0}--\eref{2.12} and~\eref{12.114}.

%           END OF SECTION~\ref{Sect5}
%<<--<--<--<--<--<--<--<--<--<--<--<--<--<--<--<--<--<--<--<--<--<--<--<--<--<

%-->-->-->-->-->-->-->-->-->-->-->-->-->-->-->-->-->-->-->-->-->-->-->-->-->>
\section
[On the momentum representation and particle interpretation]
{On the momentum representation and\\ particle interpretation}
\label{Sect6}
%           BEGINNING OF SECTION~\ref{Sect6}

    An important role in quantum field theory plays the so-called
\emph{momentum representation} (in Heisenberg
picture)~\cite{Bogolyubov&Shirkov,Bjorken&Drell-2,Itzykson&Zuber}. Its
essence is in the replacement of the field operators $\tope{\varphi}_i(x)$
with their Fourier images $\tope{\phi}_i(k)$, both connected by the
Fourier transform~($kx:=k_\mu x^\mu$)%
\footnote{~%
For brevity, we omit the inessential for us factor, equal to a power of
$2\pi$, in the r.h.s.\ of~\eref{6.1}.%
}
    \begin{equation}    \label{6.1}
\tope{\varphi}_i(x)
=
\int \e^{-\iih kx} \tope{\phi}_i(k) \Id^4 k,
    \end{equation}
and then the representation of the field equations, dynamical variables,
etc.\ in terms of $\tope{\phi}_i(k)$.

    Applying the general rule~\eref{12.114} to~\eref{6.1}, we see that
the analogue of $\tope{\phi}_i(k)$ in momentum picture is the operator
    \begin{gather}  \label{6.2}
\ope{\phi}_i(k)
:=
\e^{-\iih kx} \ope{U}(x,x_0)\circ \tope{\phi}_i(k)\circ \ope{U}^{-1}(x,x_0),
\\\intertext{which is independent of $x$, depends generally on $x_0$ and is
such that}
        \label{6.3}
\varphi_i = \int \phi_i(k) \Id^4k.
    \end{gather}

    A field theory in terms of the operators $\phi_i(k)$ will be said to
be in the \emph{momentum representation} in momentum picture.

    The Heisenberg relation~\eref{4.18b} in momentum representation,
evidently, reads
    \begin{equation}    \label{6.4}
[\tope{\phi}_i(k), \tope{P}_\mu]_{\_}  = - k_\mu \tope{\phi}_i(k)
\qquad
[\ope{\phi}_i(k), \tope{P}_\mu]_{\_}   = - k_m \ope{\phi}_i(k)
    \end{equation}
in Heisenberg and momentum picture, respectively.%
\footnote{~%
The equations~\eref{6.4} are a particular realization of a general rule,
according to which any linear combination, possibly with operator
coefficients, of $\tope{\varphi}_i(x)$ and their partial derivatives (up to a
finite order) transforms into a polynomial in $k_\mu$, the coefficients of
which are proportional to $\tope{\phi}_i(k)$. By virtue of~\eref{6.2}, the
same result holds in terms of $\ope{\phi}_i(k)$ instead of
$\tope{\phi}_i(k)$, i.e.\ in momentum picture.%
}
Consider a state vector $\tope{X}_p$ with fixed 4\ndash momentum
$p=(p_0,\dots,p_3)$, \ie for which
    \begin{equation}    \label{6.5}
\tope{P}_\mu(\tope{X}_p) = p_\mu \tope{X}_p
\qquad
  \ope{P}_\mu(\ope{X}_p) = p_\mu \ope{X}_p.
    \end{equation}
Combining these equations with~\eref{6.4}, we get
    \begin{equation}    \label{6.6}
\tope{P}_\mu\big( \tope{\phi}_i(k) (\tope{X}_p)) \bigr)
= (p_\mu + k_\mu) \tope{\phi}_i(k) (\tope{X}_p))
\qquad
\ope{P}_\mu\big( \ope{\phi}_i(k) (\ope{X}_p)) \bigr)
= (p_\mu + k_\mu) \ope{\phi}_i(k) (\ope{X}_p)).
    \end{equation}
So, the operators $\tope{\phi}_i(k)$ and $\ope{\phi}_i(k)$ increase the
state's 4\ndash momentum $p_\mu$ by $k_\mu$. If it happens that $k_0\ge0$, we
can say that these operators create a particle with 4\ndash momentum
$(\sqrt{k^2-\bk^2},\bk)$. (Notice $k^2=k_0^2+\bk^2$, $\bk:=(k^1,k^2,k^3)$,
need not to be a constant in the general case, so the mass
$m:=\frac{1}{c}\sqrt{k^2}$ is, generally, momentum\ndash dependent.) One can
introduce the creation/annihilation operators by
    \begin{equation}    \label{6.7}
\tope{\phi}_i^{\pm}(k)
:=  \begin{cases}
\tope{\phi}_i(\pm k)        &\text{for } k_0\ge0    \\
\frac{1}{2}\tope{\phi}_i(\pm k) &\text{for } k_0=0  \\
0           &\text{for } k_0<0
    \end{cases}
\qquad
\ope{\phi}_i^{\pm}(k)
:=  \begin{cases}
\ope{\phi}_i(\pm k)     &\text{for } k_0\ge0    \\
\frac{1}{2}\ope{\phi}_i(\pm k)  &\text{for } k_0=0  \\
0           &\text{for } k_0<0
    \end{cases}
\ .
    \end{equation}
In terms of them, equations~\eref{6.6} take the form
    \begin{equation}    \label{6.8}
\tope{P}_\mu\big( \tope{\phi}_i^\pm(k) (\tope{X}_p)) \bigr)
= (p_\mu \pm k_\mu) \tope{\phi}_i(k) (\tope{X}_p))
\qquad
\ope{P}_\mu\big( \ope{\phi}_i^\pm(k) (\ope{X}_p)) \bigr)
= (p_\mu \pm k_\mu) \ope{\phi}_i(k) (\ope{X}_p)).
    \end{equation}
Thus, if $k_0\ge0$, we can interpret
$\tope{\phi}_i^+(k)$ and $\ope{\phi}_i^+(k)$ (resp.\
$\tope{\phi}_i^-(k)$ and $\ope{\phi}_i^-(k)$ )
as operators creating (resp.\ annihilating) a particle with 4-\ndash momentum
$k_\mu$.

    If the relations~\eref{2.17} (resp.~\eref{2.18}) hold, similar
considerations are (resp.\ partially) valid with respect to state vectors with
fixed charge (resp.\ total angular momentum).

    As we see, the description of a quantum field theory in momentum
representation is quite similar in Heisenberg picture, via the operators
$\tope{\phi}_i(k)$, and in momentum picture, via the operators
$\ope{\phi}_i(k)$. This similarity will be investigated deeper on concrete
examples in forthcoming paper(s). The particular form of the operators
$\tope{\phi}_i(k)$ and $\ope{\phi}_i(k)$ can be found by solving the field
equations, respectively~\eref{2-2} and~\eref{12.129}, in momentum
representation, but the analysis of the so\ndash arising equations is out of
the subject of the present work.

%           END OF SECTION~\ref{Sect6}
%<<--<--<--<--<--<--<--<--<--<--<--<--<--<--<--<--<--<--<--<--<--<--<--<--<--<

%-->-->-->-->-->-->-->-->-->-->-->-->-->-->-->-->-->-->-->-->-->-->-->-->-->>
\section
[The momentum picture as 4-dimensional analogue of the Schr{\"o}dinger one]
{The momentum picture as 4-dimensional analogue of\\ the Schr{\"o}dinger one}
\label{Sect7}
%           BEGINNING OF SECTION~\ref{Sect7}

    We have introduced the momentum picture and explored some its
aspects on the base of the Heisenberg one, \ie the latter  picture was taken
as a ground on which the former one was defined and investigated; in
particular, the conditions~\eref{4.18} turn to be important from this
view\ndash point. At that point, a question arises: can the momentum picture
be defined independently and to be taken as a base from which the Heisenberg
one to be deduced? Below is presented a partial solutions of that problem
for theories in which the equations~\eref{4.18} hold.

    First of all, it should be decided which properties of the momentum
picture, considered until now, characterize it in a more or less unique way
and then they or part of them to be incorporated in a suitable (axiomatic)
definition of momentum picture. As a guiding idea, we shall follow the
understanding that the momentum picture is (or should be) a 4\ndash
dimensional analogue of the Schr{\"o}dinger picture in non\ndash relativistic
quantum mechanics. Recall,~\cite{Messiah-QM,Dirac-PQM,Fock-FQM}, the latter is
defined as a representation of quantum mechanics in which: (i)~the operators,
corresponding to the dynamical variables, are time\ndash independent;
(ii)~these operators are taken as predefined (granted) in an appropriate way;
and (iii)~the wavefunctions $\psi$ are, generally, time\ndash dependent and
satisfy the Schr{\"o}dinger equation
    \begin{equation}    \label{7.1}
\frac{\pd\psi}{\pd t} = \iih \mathcal{H}(\psi) ,
    \end{equation}
with $\mathcal{H}$ being the system's Hamiltonian acting on the system's
Hilbert space of states.  A 4\ndash dimensional generalization of
(i)\Ndash(iii), adapted for the needs of quantum field theory, will result
in an independent definition of the momentum picture. Since in that theory the
operators of the dynamical variables are constructed form the field operators
$\varphi_i$, the latter should be used for the former ones when the
generalization mentioned is carried out. Besides, the field operators satisfy
some equations, which have no analogues in quantum mechanics, which indicates
to a nontrivial generalization of item (ii) above.

    Following these ideas, we define the \emph{momentum picture} of
quantum field theory as its representation in which:\\
    \indent(a)~The field operators $\varphi_i$ are spacetime-independent,
    \begin{equation}    \label{7.2}
\pd_\mu(\varphi_i) = 0.
    \end{equation}
    \indent(b)~The state vectors $\chi$ are generally spacetime-dependent
and satisfy the following first order system of partial differential
equations
    \begin{equation}    \label{7.3}
\pd_\mu(\chi) =\iih \ope{P}_\mu(\chi) ,
    \end{equation}
where $\ope{P}_\mu$ are the components of the system's momentum operator
(constructed according to point~(c) below -- see~\eref{7.7}). If
$\chi_0\in\Hil$ and $x_0\in M$ are fixed, the system~\eref{7.3} is supposed
to have a unique solution satisfying the initial condition
    \begin{equation}    \label{7.3-1}
\chi|_{x=x_0} = \chi_0 .
    \end{equation}
    \indent(c)~If $\tope{D}(\tope{\varphi}_i,\pd_\mu\tope{\varphi}_j)$
is the density current of a dynamical variable in (ordinary) Heisenberg
picture, which is supposed to be polynomial or convergent power series in
$\tope{\varphi}_i$ and $\pd_\mu\tope{\varphi}_j$, then this quantity in
momentum picture is defined to be
    \begin{equation}    \label{7.4}
\ope{D}
= \ope{D}(\varphi_i)
:= \tope{D}\bigl( \varphi_i,\iih[\varphi_j,\ope{P}_\mu]_{\_} \bigr).
    \end{equation}
The corresponding spacetime conserved operator is defined as
    \begin{equation}    \label{7.5}
\Dsf
:= \frac{1}{c} \int_{x_0=\const}
\ope{U}^{-1}(x,x_0)\circ\ope{D}(\varphi_i)\circ\ope{U}(x,x_0)  \Id^3\bs{x} ,
    \end{equation}
where $\ope{U}(x,x_0)$ is the evolution operator
for~\eref{7.3}--\eref{7.3-1}, \ie the unique solution of the initial\ndash
value problem
    \begin{subequations}    \label{7.6}
    \begin{align}   \label{7.6a}
\frac{\pd\ope{U}(x,x_0)}{\pd x^\mu} &= \iih \ope{P}_\mu\circ\ope{U}(x,x_0)
\\          \label{7.6b}
\ope{U}(x_0,x_0) &= \id_\Hil
    \end{align}
    \end{subequations}
with $\ope{P}_\mu$ corresponding to~\eref{7.5} with the energy\ndash
momentum tensor $\ope{T}_{\mu\nu}$ for $\ope{D}$,
    \begin{equation}    \label{7.7}
\ope{P}_\mu
:= \frac{1}{c} \int_{x_0=\const}
\ope{U}^{-1(x,x_0)}\circ\ope{T}_{0\mu}(\varphi_i)\circ\ope{U}(x,x_0)
\Id^3\bs{x} .
    \end{equation}
    \indent(d)~The field operators $\varphi_i$ are solutions  of the
(algebraic) field equations, which (in the most cases) are identified with
the Euler\ndash Lagrange equations
    \begin{equation}    \label{7.8}
\Bigl\{
\frac{\pd\tope{L}(\varphi_j,y_{l\nu})} {\pd \varphi_i}
-
\iih
\Bigl[
\frac{\pd\tope{L}(\varphi_j,y_{l\nu})} {y_{i\mu}} , \ope{P}_{\mu}
\Bigr]_{\_}
\Bigr\}
\Big|_{ y_{j\nu}=\iih[\varphi_j,\ope{P}_{\nu} ]_{\_} }
= 0 ,
    \end{equation}
with $\tope{L}(\varphi_j,\iih[\varphi_j,\ope{P}_{\nu} ]_{\_})$
being the system's Lagrangian (in momentum picture, defined according
to~\eref{7.4}).

    A number or comments on the conditions (a)--(d) are in order.

    The transition from momentum to Heisenberg picture is provided by the
inversion of~\eref{12.113} and~\eref{12.113} with $\ope{U}(x,x_0)$ given
via~\eref{7.6}, \ie
    \begin{align}   \label{7.9}
\ope{X}\mapsto \tope{X}
    &= \ope{U}^{-1}(x,x_0) (\tope{X}(x))
\\          \label{7.10}
\ope{A}(x)\mapsto \tope{A}(x)
    &= \ope{U}^{-1}(x,x_0)\circ (\ope{A}(x)) \circ \ope{U}(x,x_0) .
    \end{align}
Since~\eref{7.6} implies
    \begin{equation}    \label{7.11}
\ope{H}_\mu(x,x_0) = \iih \ope{P}_\mu
    \end{equation}
for the quantities~\eref{4.4}, the replacement~\eref{4.6} is valid. In
particular, we have
    \begin{equation}    \label{7.12}
\pd_\mu\tope{\varphi}_i \mapsto
y_{j\mu}=\iih[\varphi_j,\ope{P}_{\mu} ]_{\_} \ ,
    \end{equation}
by virtue of~\eref{7.2}, which justifies the definition~\eref{7.4} and the
equation~\eref{7.8}. The Heisenberg relations~\eref{4.18b} follow from this
replacement:
\[
[\tope{\varphi}_i(x),\tope{P}_\mu]_{\_}
=
\ope{U}^{-1}(x,x_0)
\circ [\ope{\varphi}_i,\ope{P}_\mu]_{\_} \circ
\ope{U}(x,x_0)
= \ih \pd_\mu(\tope{\varphi}_i).
\]

    Since the integrability conditions for~\eref{7.3} are
    \begin{align*}
0
& =
\pd_\nu\circ\pd_\mu (\chi) - \pd_\mu\circ\pd_\nu (\chi)
  =
\iih \bigl\{ \pd_\nu(\ope{P}_\mu(\chi)) - \pd_\mu(\ope{P}_\nu(\chi)) \bigr\}
\\
& =
\iih\bigl\{
(\pd_\nu(\ope{P}_\mu)-\pd_\mu(\ope{P}_\nu))(\chi) +
\ope{P}_\mu(\pd_\nu(\chi)) - \ope{P}_\nu(\pd_\mu(\chi))
\bigr\} ,
    \end{align*}
where~\eref{7.3} was applied, the existence of a unique solution
of~\eref{7.3}--\eref{7.3-1} implies (use~\eref{7.2} again;
\cf footnote~\ref{fn:momentum-curvature})
    \begin{equation}    \label{7.13}
\pd_\nu(\ope{P}_\mu)-\pd_\mu(\ope{P}_\nu)) +
\iih[\ope{P}_\mu,\ope{P}_\nu]_{\_}
=0 .
    \end{equation}
As $\pd_\nu\tope{P}_\mu=0$, due to the conservation of $\tope{P}_\mu$, the
replacement~\eref{4.3}, with $\ope{P}_\nu$ for $\ope{A}(x)$, together
with~\eref{7.11} entails
$ \pd_\mu(\ope{P}_\nu)) + \iih[\ope{P}_\nu,\ope{P}_\mu]_{\_}=0 $,
which, when inserted into~\eref{7.13}, gives
    \begin{equation}    \label{7.14}
\pd_\nu(\ope{P}_\mu) = 0 .
    \end{equation}
The substitution of~\eref{7.14} into~\eref{7.13} results in
    \begin{equation}    \label{7.15}
[\ope{P}_\mu,\ope{P}_\nu]_{\_} = 0 ,
    \end{equation}
which immediately implies~\eref{4.18a}.

    As a result of~\eref{7.15} and~\eref{7.6}, we obtain
    \begin{equation}    \label{7.16}
\ope{U}(x,x_0) = \e^{\iih(x^\mu-x_0^\mu)\ope{P}_\mu},
    \end{equation}
so that
    \begin{equation}    \label{7.17}
   [\ope{U}(x,x_0),\ope{P}_\mu]_{\_}
=  [\ope{U}^{-1}(x,x_0),\ope{P}_\mu]_{\_}
= 0
    \end{equation}
and, consequently
    \begin{equation}    \label{7.18}
\tope{P}_\mu=\ope{U}^{-1}(x,x_0)\circ\ope{P}_\mu\circ\ope{U}(x,x_0)
=
\ope{P}_\mu ,
    \end{equation}
which implies the coincidence of the evolution operators given
by~\eref{12.112} and~\eref{7.6}. The last conclusion leads to the
identification of the momentum picture defined via the conditions
(a)\Ndash(d) above and by~\eref{12.113}, \eref{12.114} and~\eref{4.18} in
Sect.~\ref{Sect4}.

    What regards the conditions (c) and (d) in the definition of the
momentum picture, they have no analogues in quantum mechanics. Indeed,
equations~\eref{7.4}--\eref{7.8} form a closed system for determination of
the field operators (via the so\ndash called creation and annihilation
operators) and, correspondingly, they provide a method for obtaining explicit
forms of the dynamical variables (via the same operators). On the contrary,
in quantum mechanics there is no procedure for determination of the operators
of the dynamical variables and they are defined by reasons external to this
theory.

    Thus, we see that a straightforward generalization of the
Schr{\"o}dinger picture in quantum mechanics to the momentum picture in quantum
field theory (expressed first of all by~\eref{7.2} and~\eref{7.3}) is
possible if and only if the equations~\eref{4.18} are valid for the system
considered.

%           END OF SECTION~\ref{Sect7}
%<<--<--<--<--<--<--<--<--<--<--<--<--<--<--<--<--<--<--<--<--<--<--<--<--<--<

%-->-->-->-->-->-->-->-->-->-->-->-->-->-->-->-->-->-->-->-->-->-->-->-->-->>
\section {Conclusion}
\label{Conclusion}
%           BEGINNING OF SECTION~\ref{Conclusion}

    In the present paper, we have summarized, analyzed and developed the
momentum picture of motion in (Lagrangian) quantum field theory, introduced
in~\cite{bp-QFT-pictures}. As it was shown, this picture is (expected to be)
useful when the conditions~\eref{4.18} are valid in (or compatible with) the
theory one investigates. If this is the case, the momentum picture has
properties that allow  one to call it a `4\ndash dimensional Schr{\"o}dinger
picture' as the field operators (and functions which are polynomial in them
and their derivatives) in it became spacetime\ndash constant operators and
the state vectors  have a simple, exponential, dependence on the spacetime
coordinates/points. This situation is similar to the one in quantum mechanics
in Schr{\"o}dinger picture, when time\ndash independent Hamiltonians are
employed~\cite{Messiah-QM}, the time replacing the spacetime coordinates in
our case.

    As we said in Sect~\ref{Sect4}, there are evidences that the
conditions~\eref{4.18} should be a part of the basic postulates of quantum
field theory (see also~\cite[\S~68]{Bjorken&Drell-2}). In the ordinary field
theory, based on the Lagrangian formalism to which (anti)commutation
relations are added as additional
conditions~\cite{Bogolyubov&Shirkov,Bjorken&Drell,Itzykson&Zuber}, the
validity of~\eref{4.18} is questionable and should be checked for any
particular Lagrangian~\cite{Bjorken&Drell-2}. The cause for this situation
lies in the fact that~\eref{4.18} and the (anti)commutation relations are
additional to the Lagrangian formalism and their compatibility is a problem
whose solution is not obvious. The solution of that problem is known to be
positive for a lot of particular Lagrangians~\cite{Bjorken&Drell-2}, but, in
the general case, it seems not to be explored. For these reasons, one may try
to `invert' the situation, \ie to consider a Lagrangian formalism, to which
the conditions~\eref{4.18} are imposed as subsidiary restrictions, and then
to try to find (anti)commutation relations that are consistent with the
so\ndash arising scheme. We intend to realize this program in forthcoming
papers, in which it will be demonstrated that the proposed method
reproduces most of the known results, reveals ways for their generalizations
at different stages of the theory, and also gives new results, such as a
(second) quantization of electromagnetic field in Lorentz gauge, imposed
directly on the field's operator\ndash valued potentials, and a `natural'
derivation of the paracommutation relations.

%           END OF SECTION~\ref{Conclusion}
%<<--<--<--<--<--<--<--<--<--<--<--<--<--<--<--<--<--<--<--<--<--<--<--<--<--<

%-->-->-->-->-->-->-->-->-->-->-->-->-->-->-->-->-->-->-->-->-->-->-->-->-->>
%\section*{Acknowledgments}
%           BEGINNING OF SECTION~\ref{Acknowledgments}

%           END OF SECTION~\ref{Acknowledgments}
%<<--<--<--<--<--<--<--<--<--<--<--<--<--<--<--<--<--<--<--<--<--<--<--<--<--<

%-->-->-->-->-->-->-->-->-->-->-->-->-->-->-->-->-->-->-->-->-->-->-->-->-->>
%           BEGINNING OF BIBLIOGRAPHY
\addcontentsline{toc}{section}{References}
\bibliography{bozhopub,bozhoref}

\begin{thebibliography}{10}

\bibitem{bp-QFT-pictures}
Bozhidar~Z. Iliev.
\newblock Pictures and equations of motion in {Lagrangian} quantum field
  theory.
\newblock In Frank Columbus, editor, {\em Progress in Mathematical Physics},
  pages ??--?? Nova Science Publishers, Inc., Suite, 2004.
\newblock To appear.\\ http://www.arXiv.org e-Print archive, E-print No.\
  hep-th/0302002, February 2003.

\bibitem{Messiah-QM}
A.~M.~L. Messiah.
\newblock {\em Quantum mechanics}, volume I and II.
\newblock Interscience, New York, 1958.
\newblock Russian translation: Nauka, Moscow, 1978 (vol.~I) and 1979 (vol.~II).

\bibitem{Bogolyubov&Shirkov}
N.~N. Bogolyubov and D.~V. Shirkov.
\newblock {\em Introduction to the theory of quantized fields}.
\newblock Nauka, Moscow, third edition, 1976.
\newblock In Russian. English translation: Wiley, New York, 1980.

\bibitem{Bjorken&Drell}
J.~D. Bjorken and S.~D. Drell.
\newblock {\em Relativistic quantum mechanics}, volume 1 and 2.
\newblock McGraw-Hill Book Company, New York, 1964, 1965.
\newblock Russian translation: Nauka, Moscow, 1978.

\bibitem{Roman-QFT}
Paul Roman.
\newblock {\em Introduction to quantum field theory}.
\newblock John Wiley\&Sons, Inc., New York-London-Sydney-Toronto, 1969.

\bibitem{Ryder-QFT}
Lewis~H. Ryder.
\newblock {\em Quantum field theory}.
\newblock Cambridge Univ.\ Press, Cambridge, 1985.
\newblock Russian translation: Mir, Moscow, 1987.

\bibitem{Akhiezer&Berestetskii}
A.~I. Akhiezer and V.~B. Berestetskii.
\newblock {\em Quantum electrodynamics}.
\newblock Nauka, Moscow, 1969.
\newblock In Russian. English translation: Authorized English ed., rev. and
  enl. by the author, Translated from the 2d Russian ed. by G.M.~Volkoff, New
  York, Interscience Publishers, 1965. Other English translations: New York,
  Consultants Bureau, 1957; London, Oldbourne Press, 1964, 1962.

\bibitem{Ramond-FT}
Pierre Ramond.
\newblock {\em Field theory: a modern primer}, volume~51 of {\em Frontiers in
  physics}.
\newblock Reading, MA Benjamin-Cummings, London-Amsterdam-Don Mills,
  Ontario-Sidney-Tokio, 1 edition, 1981.
\newblock 2nd rev.\ print, Frontiers in physics vol.~74, Adison Wesley Publ.\
  Co., Redwood city, CA, 1989; Russian translation from the first ed.: Moscow,
  Mir 1984.

\bibitem{Bogolyubov&et_al.-AxQFT}
N.~N. Bogolubov, A.~A. Logunov, and I.~T. Todorov.
\newblock {\em Introduction to axiomatic quantum field theory}.
\newblock W.~A. Benjamin, Inc., London, 1975.
\newblock Translation from Russian: Nauka, Moscow, 1969.

\bibitem{Bogolyubov&et_al.-QFT}
N.~N. Bogolubov, A.~A. Logunov, A.~I. Oksak, and I.~T. Todorov.
\newblock {\em General principles of quantum field theory}.
\newblock Nauka, Moscow, 1987.
\newblock In Russian. English translation: Kluwer Academic Publishers,
  Dordrecht, 1989.

\bibitem{Itzykson&Zuber}
C.~Itzykson and J.-B. Zuber.
\newblock {\em Quantum field theory}.
\newblock McGraw-Hill Book Company, New York, 1980.
\newblock Russian translation (in two volumes): Mir, Moscow, 1984.

\bibitem{bp-QFT-action-principle}
Bozhidar~Z. Iliev.
\newblock On operator differentiation in the action principle in quantum field
  theory.
\newblock In Stancho Dimiev and Kouei Sekigava, editors, {\em Proceedings of
  the 6th International Workshop on Complex Structures and Vector Fields, 3--6
  September 2002, St.\ Knstantin resort (near Varna), Bulgaria}, ``Trends in
  Complex Analysis, Differential Geometry and Mathematical Physics'', pages
  76--107. World Scientific, New Jersey-London-Singapore-Hong Kong, 2003.
\newblock \\ http://www.arXiv.org e-Print archive, E-print No.\ hep-th/0204003,
  April 2002.

\bibitem{Bjorken&Drell-2}
J.~D. Bjorken and S.~D. Drell.
\newblock {\em Relativistic quantum fields}, volume~2.
\newblock McGraw-Hill Book Company, New York, 1965.
\newblock Russian translation: Nauka, Moscow, 1978.

\bibitem{Ohnuki&Kamefuchi}
Y.~Ohnuki and S.~Kamefuchi.
\newblock {\em Quantum field theory and parafields}.
\newblock University of Tokyo Press, Tokyo, 1982.

\bibitem{Dirac-PQM}
P.~A.~M. Dirac.
\newblock {\em The principles of quantum mechanics}.
\newblock Oxford at the Clarendon Press, Oxford, fourth edition, 1958.
\newblock Russian translation in: P.~Dirac, Principles of quantum mechanics,
  Moscow, Nauka, 1979.

\bibitem{Fock-FQM}
V.~A. Fock.
\newblock {\em Fundamentals of quantum mechanics}.
\newblock Mir Publishers, Moscow, 1978.
\newblock Russian edition: Nauka, Moscow, 1976.

\end{thebibliography}
\bibliographystyle{unsrt}
\addcontentsline{toc}{subsubsection}{This article ends at page}
%\addtocontents{toc}{}
%           END OF BIBLIOGRAPHY
%<<--<--<--<--<--<--<--<--<--<--<--<--<--<--<--<--<--<--<--<--<--<--<--<--<--<

\end{document}